\documentclass[conference]{IEEEtran}
\usepackage[style=ieee, backend=biber, isbn=false]{biblatex}

\usepackage{amsmath,amssymb,amsfonts}
\usepackage{algorithmic}
\usepackage{graphicx}
\usepackage{textcomp}
\usepackage{xcolor}
\usepackage{siunitx}
\usepackage{comment}
\def\BibTeX{{\rm B\kern-.05em{\sc i\kern-.025em b}\kern-.08em T\kern-.1667em\lower.7ex\hbox{E}\kern-.125emX}}

\usepackage{import}

\usepackage[utf8]{luainputenc}

\usepackage[T1]{fontenc}

\usepackage{graphicx}

\usepackage[english]{babel}
\addto\captionsenglish{}
\addto\captionsenglish{}
\usepackage{csquotes}

\usepackage{newtxtext}
\usepackage{amsthm}
\usepackage[slantedGreek]{newtxmath}
\usepackage[OMLmathsfit]{isomath}
\DeclareMathAlphabet{\mathbfsf}{\encodingdefault}{\sfdefault}{bx}{n}
\usepackage{bm}
\usepackage{envmath}
\usepackage{mathtools}
\usepackage{commath}
\usepackage{siunitx}

\usepackage[caption=false,font=footnotesize]{subfig}

\usepackage{booktabs}
\usepackage{footmisc}  

\usepackage{url}

\theoremstyle{definition}

\theoremstyle{plain}

\theoremstyle{remark}

\usepackage{lineno}
\modulolinenumbers[5]
\usepackage{umoline}

\usepackage{pgfplots}
\usepackage{pgfplotstable}
\pgfplotsset{compat=newest}
\pgfplotsset{plot coordinates/math parser=false}
\newlength\figureheight
\newlength\figurewidth
\pgfplotsset{every axis plot/.append style={line width=1.5pt},
    legend style={font=\footnotesize, 
        text height=1.0ex,
        draw=black,
        fill=white,
        legend cell align=left}}

\usepackage{hyperref} 
\usepackage[english]{cleveref}

\Crefname{defn}{definition}{definitions}
\Crefname{defn}{Definition}{Definitions}

\Crefname{asm}{assumption}{assumptions}
\Crefname{asm}{Assumption}{Assumptions}

\crefname{lem}{lemma}{lemmas} 
\Crefname{lem}{Lemma}{Lemmas}

\crefname{prop}{proposition}{propositions} 
\Crefname{prop}{Proposition}{Propositions}

\crefname{thm}{theorem}{theorms} 
\Crefname{thm}{Theorem}{Theorms}

\crefname{cor}{corollary}{corollaries}
\Crefname{cor}{Corollary}{Corollaries}
\newcounter{subequation}
\newlength\mtabskip\mtabskip=-1.25cm

\def\mtabLong{long}

\newcommand{\mr}{\mathrm}

\newcommand{\veg}[1]{\bm{#1}}     
\newcommand{\mat}[1]{\mathsfbfit{#1}} 
\newcommand{\op}[1]{\mathcal{#1}} 
\newcommand{\vecop}[1]{\bm{\mathcal{#1}}} 

\newcommand\restr[2]{{
        \left.\kern-\nulldelimiterspace 
        #1 
        \vphantom{|} 
        \right|_{#2} 
}}

\newcommand\rst[3]{{
        \left.\kern-\nulldelimiterspace 
        #1 
        \vphantom{|} 
        \right|_{#2}^{#3} 
}}

\usepackage{acro}

\DeclareAcronym{DG}
{
    short = DG ,
    long = discontinuous Galerkin
}

\DeclareAcronym{ACA}
{
    short = ACA ,
    long = adaptive cross approximation
}

\DeclareAcronym{EFIE}
{
    short =  EFIE ,
    long = electric field integral equation
}

\DeclareAcronym{MFIE}
{
    short =  MFIE ,
    long = magnetic field integral equation
}

\DeclareAcronym{CFIE}
{
    short =  CFIE ,
    long = combined field integral equation
}

\DeclareAcronym{MUIE}
{
    short =  MUIE ,
    long = Müller integral equation
}

\DeclareAcronym{PMCHWT}
{
    short =  PMCHWT ,
    long = Poggio-Miller-Chang-Harrington-Wu-Tsai integral equation
}

\DeclareAcronym{SPD}
{
    short =  SPD ,
    long = {symmetric, positive definite}
}

\DeclareAcronym{SPSD}
{
    short =  SPD ,
    long = {symmetric, positive semi-definite}
}

\DeclareAcronym{PEC}
{
    short =  PEC ,
    long = perfectly electrically conducting
}

\DeclareAcronym{RWG}
{
    short = RWG ,
    long = Rao-Wilton-Glisson
} 

\DeclareAcronym{BC}
{
    short = BC ,
    long = Buffa-Christiansen
}

\DeclareAcronym{SVD}
{
    short = SVD ,
    long = singular value decomposition
}

\DeclareAcronym{CG}
{
    short = CG ,
    long = conjugate gradient
} 

\DeclareAcronym{PCG}
{
    short = PCG ,
    long = preconditioned conjugate gradient
} 

\DeclareAcronym{CGS}
{
    short = CGS ,
    long = conjugate gradient squared
}

\DeclareAcronym{CMP}
{
    short = CMP ,
    long = Calderón multiplicative preconditioner
} 

\DeclareAcronym{RFCMP}
{
    short = RF-CMP ,
    long = refinement-free Calderón multiplicative preconditioner
} 

\DeclareAcronym{HPD}
{
    short = HPD ,
    long = {Hermitian, positive definite}
} 

\DeclareAcronym{RHS}
{
    short = RHS ,
    long = {right-hand side}
}

\DeclareAcronym{PW}
{
    short = PW ,
    long = {plane wave}
} 

\DeclareAcronym{HD}
{
    short = HD ,
    long = {Hertzian dipole}
} 

\DeclareAcronym{FF}
{
    short = FF ,
    long = {far-field}
} 

\DeclareAcronym{NF}
{
    short = NF ,
    long = {near-field}
}  

\newcolumntype {n}{c}
\newcolumntype {N}{>{\small}c}
\newcolumntype {L}{>{\small}l}
\newcolumntype {F}{>{\footnotesize}c}
\newcolumntype {v}[1]{>{\raggedright \hspace {0pt}} p {#1}}
\newcolumntype {V}[1]{>{\small \raggedright \hspace {0pt}} p {#1}}
\newcolumntype{d}[1]{>{\DC@{.}{.}{#1}}c<{\DC@end}}

\newcolumntype{R}[1]{%
    >{\begin{turn}{90}\begin{minipage}{#1}\small\raggedright\hspace{0pt}}l%
            <{\end{minipage}\end{turn}}%
}

\NewDocumentCommand{\TA}{o}{
    \IfNoValueTF {#1} {%
        \vecop T_{\kern-2pt\mr{A}}
    }
    {
        \vecop T_{\kern-2pt\mr{A},#1}
    }
}

\NewDocumentCommand{\TPhi}{o}{
    \IfNoValueTF {#1} {%
        \vecop T_{\kern-2pt\Phiup}
    }
    {
        \vecop T_{\kern-2pt\Phiup,#1}
    }
}

\NewDocumentCommand{\matTA}{o}{
    \IfNoValueTF {#1} {%
        \mat T_\mr{A}   
        }
    {
        \mat T_{\mr{A},#1}
    }
}

\NewDocumentCommand{\matTPhi}{o}{
    \IfNoValueTF {#1} {%
        \mat T_\Phiup   
        }
    {
        \mat T_{\Phiup,#1}
    }
}

\NewDocumentCommand{\MSL}{o}{
    \IfNoValueTF {#1} {%
        \veg \Psi_\mr{SL}
        }
    {
        \veg \Psi_{\mr{SL},#1}
    }
}

\NewDocumentCommand{\MDL}{o}{
    \IfNoValueTF {#1} {%
        \veg \Psi_\mr{DL}
        }
    {
        \veg \Psi_{\mr{DL},#1}
    }
}

\NewDocumentCommand{\PA}{o}{
    \IfNoValueTF {#1} {%
        \veg \Psi_\mr{A}
        }
    {
        \veg \Psi_{\mr{A},#1}
    }
}

\NewDocumentCommand{\PPhi}{o}{
    \IfNoValueTF {#1} {%
        \veg \Psi_{\Phiup}
        }
    {
        \veg \Psi_{\Phiup,#1}
    }
}

\makeatletter

\newcounter{authr}
\newcommand{\authr}[2][]{
   \stepcounter{authr}
   \@namedef{authr@\theauthr}{#2}
   \@namedef{authrlabel@\theauthr}{#1}
}

\newcounter{address}
\newcommand{\address}[2][]{
   \stepcounter{address}
   \@namedef{address@\theaddress}{#2}
   \@namedef{addresslabel@\theaddress}{#1}
}

\newcommand{\alsep}{and}

\def\newmaketitle{\par%
  \begingroup%
  \normalfont%
  \def\thefootnote{}
  \def\footnotemark{}
  \let\@makefnmark\relax
  \footnotesize
  \footnotesep 0.7\baselineskip
  \normalsize%
  \twocolumn[\thenewmaketitle\@IEEEaftertitletext]%
  \if@IEEEusingpubid
     \enlargethispage{-\@IEEEpubidpullup}%
  \fi
  \endgroup
  \setcounter{footnote}{0}\let\maketitle\relax\let\@maketitle\relax
  \gdef\@thanks{}%
  \let\thanks\relax}

\def\thenewmaketitle{
  \newpage
  \begin{center}%
    \vskip0.2em{\Huge\@IEEEcompsoconly{\sffamily}\@IEEEcompsocconfonly{\normalfont\normalsize\vskip 2\@IEEEnormalsizeunitybaselineskip
   \bfseries\large}\@title\par}\vskip1.0em\par%
    \vspace{1ex}
    \newcounter{c@authr}
    \newcounter{c@tmp}
    \ifthenelse{\value{authr}=2}{%
      \newcommand{\liand}{ and }}{%
      \newcommand{\liand}{, and }}
    \ifthenelse{\value{address}<2}{%
      \@nameuse{authr@1}%
      \stepcounter{c@authr}%
      \whiledo{\value{c@authr}<\value{authr}}{%
        \setcounter{c@tmp}{\value{authr}}%
        \addtocounter{c@tmp}{-\value{c@authr}}%
        \ifthenelse{\value{c@tmp}=1}{%
          \renewcommand{\alsep}{\liand}}{\renewcommand{\alsep}{, }}%
        \stepcounter{c@authr}\alsep \@nameuse{authr@\thec@authr}}\\%
    }
    {
      \@nameuse{authr@1}${}^{(\ref{\@nameuse{authrlabel@1}})}$%
      \stepcounter{c@authr}%
      \whiledo{\value{c@authr}<\value{authr}}{%
      \setcounter{c@tmp}{\value{authr}}%
      \addtocounter{c@tmp}{-\value{c@authr}}%
      \ifthenelse{\value{c@tmp}=1}{%
        \renewcommand{\alsep}{\liand}}{\renewcommand{\alsep}{, }}%
      \stepcounter{c@authr}\alsep \@nameuse{authr@\thec@authr}%
        ${}^{(\ref{\@nameuse{authrlabel@\thec@authr}})}$%
      }
    }
    \vspace{0.2ex}

    \ifthenelse{\value{address}>0}{%
      \ifthenelse{\value{address}=1}{
        {\@nameuse{address@1}}
      }
      {
        \newcounter{c@address}

        \begin{center}
        \whiledo{\value{c@address}<\value{address}}
        {
          \refstepcounter{c@address}
            ${}^{(\thec@address)}$\,%
              \label{\@nameuse{addresslabel@\thec@address}}%
              \@nameuse{address@\thec@address}\\ %
        }
        \end{center}
      } 
    }
    {
      \relax
    }
  \end{center}
}

\makeatother

\title{On a High-Frequency Analysis of Some Relevant Integral Equations in Electromagnetics
}

\authr[org1]{Viviana Giunzioni}
\authr[org2]{Adrien Merlini}
\authr[org1]{Francesco P. Andriulli}

\address[org1]{Department of Electronics and Telecommunications, Politecnico di Torino, 10129 Turin, Italy}
\address[org2]{Microwaves Department, IMT Atlantique, 29238 Brest, France}

\begin{document}
\newmaketitle

\begin{abstract}
In this contribution we analyze the spectral properties of some commonly used boundary integral operators in computational electromagnetics and of their discrete counterparts, highlighting   peculiar features of their spectra. In particular, a comparison with the eigenvalues of the continuous operators will be presented that highlights deviations in the high frequency regime and impacts, in a peculiar way, the accuracy of the numerical solutions of  each formulation.
A study and a proactive analysis of numerical results from standard boundary element solvers and the predictions from the theoretical analysis will corroborate the analytical framework employed and the validity of our observations.
\end{abstract}

\begin{IEEEkeywords}
Integral operators, high-frequency, spectral analysis, accuracy of the Boundary Element Method
\end{IEEEkeywords}

\section{Introduction}
Boundary integral equations are well-established in computational electromagnetics. Numerically solved by boundary element methods (BEMs), they provide a full-wave solution to the Maxwell's system. Their accuracy and adherence to the physics offer an advantage in terms of prediction power with respect to approximate methods, such as ray tracing or geometrical optics, useful at high frequency.

When modeling the two-dimensional electromagnetic scattering from a perfectly conducting metallic obstacle, the electric and magnetic field integral equations (EFIE / MFIE) act as building blocks for other formulations, such as the combined field integral equation (CFIE) and preconditioned versions of it, freed from spurious resonances.

In this contribution, we analyze the above-mentioned integral formulations applied to infinitely extended cylinders in the high-frequency regime, characterized by a mesh discretization density increasing proportionally with the wavenumber of the impinging fields. This condition is equivalent to fixing a certain number of degrees of freedom per wavelenght. After introducing the required formalism in Section~\ref{sec:formalism}, we will provide an analysis of how the discretization process causes the spectra of BEM matrices and the spectra of the continuous operators from which they derive to differ. This analysis along with the effect of the discretization on the solution accuracy will be delineated in  Section~\ref{sec:novelty}. The numerical results proposed in Section~\ref{sec:numerical} will illustrate the theoretical developments. 

\section{Formalism}
\label{sec:formalism}
Consider the time harmonic electromagnetic scattering from a perfect electrically conducting (PEC) cylinder indefinitely extended along the longitudinal direction $\hat{\veg z}$. Let $\Omega$ be the open set modeling the transversal cross-section of a cylinder of radius $a$ and of axis along $\hat{z}$ and $\Gamma\coloneq\partial \Omega$ be its two-dimensional circular contour. The exterior space $\mathbb{R}^2\backslash\Omega$ is characterized by its impedance $\eta = \sqrt{\mu/\epsilon}$ and the corresponding wavenumber $k = \omega \sqrt{\mu\epsilon}$. We define the single layer, double layer, adjoint double layer, and hypersingular operators respectively as
\begin{align}
    \op S^k f(\veg \rho) &\coloneqq k \int_\Gamma G^k (\veg \rho,\veg \rho') f(\veg \rho') d \veg \rho'\,,\\
    \op D^k f(\veg \rho) &\coloneqq \int_\Gamma \frac{\partial}{\partial n'} G^k (\veg \rho,\veg \rho') f(\veg \rho') d \veg \rho'\,,\\
    \op D^{*k} f(\veg \rho) &\coloneqq \int_\Gamma \frac{\partial}{\partial n} G^k (\veg \rho,\veg \rho') f(\veg \rho') d \veg \rho'\,,\\
    \op N^k f(\veg \rho) &\coloneqq - \frac{1}{k}\frac{\partial}{\partial n} \int_\Gamma \frac{\partial}{\partial n'} G^k (\veg \rho,\veg \rho') f(\veg \rho') d \veg \rho'\,, 
\end{align}
where $G^k$ is the two-dimensional free-space Green's function, $G^k(\veg \rho, \veg \rho') \coloneqq -\frac{\mathrm{j}}{4} H_0^{(2)} (k|\veg \rho- \veg \rho'|)$.
These are the building blocks for the electric and magnetic field integral equations that relate the longitudinal and transversal electric current densities $J_z$ and $J_t$ and the impinging electromagnetic fields $(E_z,H_t)$ and $(E_t,H_z)$. In the transverse magnetic (TM) polarization, they read
\begin{align}
    \op S^k (J_z) (\veg \rho) &= \frac{E_z(\veg \rho)}{\mathrm{j} \eta} \,,\\
    \left(\frac{1}{2} \op I +  \op D^{*k} \right) (J_z) (\veg \rho) &= H_t(\veg \rho) \,,
\end{align}
while in the transverse electric (TE) polarization they read
\begin{align}
     \op N^k (J_t) (\veg \rho) &= -\frac{ E_t(\veg \rho)}{\mathrm{j}\eta}\,, \\
    \left(\frac{1}{2} \op I -  \op D^k \right) (J_t) (\veg \rho) &= - H_z(\veg \rho)\,.
\end{align}
The EFIE and MFIE in the high-frequency regime are plagued by ill conditioning, known as the high-frequency breakdown, as well as  by spurious resonances. Both these issues can be addressed by combining and preconditioning these equations to form the Calder\'{o}n combined field integral equation, referred in the following as CCFIE, that reads for TM and TE polarizations respectively
\begin{align}
    &\left[{\op N^{\tilde{k}}} \op S^k  + \left(\frac{1}{2} \op I -  {\op D}^{*\tilde{k}} \right)\left(\frac{1}{2} \op I +  \op D^{*k} \right)\right] (J_z) (\veg \rho) =\nonumber\\ 
    &\quad\quad\quad\quad\quad\quad \frac{\op N^{\tilde{k}}}{\mathrm{j}\eta}E_z(\veg \rho)+\left(\frac{1}{2} \op I -  {\op D}^{*\tilde{k}} \right)H_t(\veg \rho)\,,\\
    &\left[{\op S^{\tilde{k}}} \op N^k  + \left(\frac{1}{2} \op I +  {\op D^{\tilde{k}}} \right)\left(\frac{1}{2} \op I -  \op D^k \right)\right] (J_t) (\veg \rho) =\nonumber\\
    &\quad\quad\quad\quad\quad\quad-\frac{\op S^{\tilde{k}}}{\mathrm{j}\eta}E_t(\veg \rho)-\left(\frac{1}{2} \op I +  {\op D^{\tilde{k}}} \right)H_z(\veg \rho)\,,
\end{align}
where $\tilde{k} \coloneqq k-\mathrm{j}0.4k^{1/3}a^{-2/3}$ \cite{antoine2006improved}.

By denoting $\lambda_q^{\op O}$ the eigenvalues of a placeholder continuous operator $\op O$, because $\Gamma$ is a circle, one can show that
\begin{align}
    \lambda_q^{\op S^k} &= -\frac{\mathrm{j} k \pi a}{2} J_{q}(k a) H_{q}^{(2)}(k a)\,,
    \label{eqn:lamS}\\
    \lambda_q^{\op D^k} &= \lambda_q^{\op D^{*k}} = -\frac{\mathrm{j} k \pi a}{4} \left[J_{q}(k a) H_{q}^{(2)}(k a) \right]'\,,
    \label{eqn:lamD}\\
    \lambda_q^{\op N^k} &= \frac{\mathrm{j} k \pi a}{2} J'_{q}(k a) H_{q}'^{(2)}(k a)\,.
    \label{eqn:lamN}
\end{align}
When discretizing one of $\op{O}$ with the BEM with test functions $t_m$ and source basis functions $f_n$, defined on a uniform mesh of $\Gamma$ characterized by $N$ mesh elements of length $h$, the element $(m,n)$ of the resulting matrix $\mat{O}$ is 
\begin{equation}
    \mat{O}_{mn} = \frac{1}{h}\int_\Gamma \mathrm{d}s \, t_m(s) \,  \left(\op O f_n\right)(s)\,,
\end{equation}
and the eigenvalues of the matrix are in the form \cite{warnick2008numerical}
\begin{equation}
    \hat{\lambda}_q^{\mat O} = \sum_{s=-\infty}^{\infty}\lambda^{\op O}_{(q+sN)}T_{-(q+sN)}F_{(q+sN)}\,,
\end{equation}
where $T_q$ and $F_q$ represents the $q-$th Fourier coefficient of the test and source basis functions.
From this, a spectral relative error can be defined as
\begin{equation}
    E_q^{\op O} \coloneq \frac{\hat{\lambda}_q^{\mat O}-{\lambda}_q^{\op O}}{{\lambda}_q^{\op O}} = E_q^P+E_q^{A,\op O}\,,
\end{equation}
where
\begin{align}
    E_q^P &\coloneq T_{-q}F_q-1\,,\\
    E_q^{A,\op O} &\coloneq \frac{1}{\lambda_q^{\op O}}\sum_{s\ne 0}\lambda^{\op O}_{(q+sN)}T_{-(q+sN)}F_{(q+sN)}
\end{align}
represent a projection and aliasing error contributions \cite{warnick2008numerical}.

\section{High-Frequency Analysis}
\label{sec:novelty}
In this contribution, we propose a high-frequency analysis of the boundary integral equations described above in terms of spectral and current error, i.e., we analyze the accuracy achievable by BEM formulations in the presence of discretization-related errors (neglecting all other possible sources of error) when increasing the frequency and the number of unknowns proportionally.

\subsection{Spectral Error}
\label{sec:spectralerror}
First, we study the relative difference between the eigenvalues of the continuous and discrete operators $E_q$ for indices $q<(ka)$ (hyperbolic region), $q \simeq (ka)$ (transition region), and $q > (ka)$ (elliptic region). 
The analyses leverage on different asymptotic expansions of the special functions in equations \eqref{eqn:lamS}, \eqref{eqn:lamD}, \eqref{eqn:lamN} depending on the regime. In particular we use large argument expansions  (\cite[Section 9.2]{abramowitz1964handbook}) in the hyperbolic region and large order expansions (\cite[Section 9.3]{abramowitz1964handbook}) in the transition and elliptic regions.
By applying these expansions, we notice that the modulus of the eigenvalues in the transition region increases as $(ka)^{1/3}$ and decreases as $(ka)^{-1/3}$, respectively, for $\op S$ and $\op N$, while the remaining part of the spectra has a constant behavior. This corresponds to a decreasing as $(ka)^{-1/3}$ and increasing as $(ka)^{1/3}$ contribution of $|E_q^{A,\op S}|$ and $|E_q^{A,\op N}|$ for indices $q\simeq (ka)$. By following the same reasoning, $|E_q^{A,\op D}|$ decays as $(ka)^{-1}$ in the hyperbolic and transition regions (away from resonances), while the discretization of the identity operator (i.e. the gram matrix) is characterized by a constant in frequency aliasing contribution. The magnetic field integral operator (MFIO), given by the sum of two commuting operators, is characterized by the relative spectral error
\begin{equation}
    E_q^{\text{TM/TE-MFIO}^k}=E_q^P + \frac{\frac{1}{2}E_q^{A,\op{I}} \pm \lambda_q^{\op{D}^k}E_q^{A,\op{D}^k}}{\frac{1}{2}\pm\lambda_q^{\op{D}^k}}
\end{equation}
dominated by the identity contribution in the aliasing term and hence constant in the high-frequency limit.
The Calder\'{o}n combined field integral operator (CCFIO), sum of the Calder\'{o}n electric and Calder\'{o}n magnetic field integral operators (CEFIO and CMFIO), is affected by the error
\begin{equation}
    E_q^{\text{CCFIO}^k}=\frac{\lambda_q^{\text{CEFIO}^k}E_q^{\text{CEFIO}^k}+\lambda_q^{\text{CMFIO}^k}E_q^{\text{CMFIO}^k}}{\lambda_q^{\text{CEFIO}^k}+\lambda_q^{\text{CMFIO}^k}}\,,
\end{equation}
where
\begin{align}
    E_q^{\text{TM/TE-CEFIO}^k} &=\frac{(1+E_q^{\op N^{\tilde{k}/k}})(1+E_q^{\op S^{k/\tilde{k}}})}{(1+E_q^{\op I})}-1\,,\\
     E_q^{\text{TM/TE-CMFIO}^k} &=\frac{(1+E_q^{\text{TE-MFIO}^{\tilde{k}/k}})(1+E_q^{\text{TM-MFIO}^{k/\tilde{k}}})}{(1+E_q^{\op I})}-1\,.
\end{align}
The presence of the hypersingular operator in both the TM and TE formulations causes an increase of the spectral relative error in the transition region as $k^{1/3}$ of the CCFIO.

\subsection{Current Error}
The relative error between the currents from the discrete ($\hat{J}$) and continuous ($J$) formulations evaluated at the mesh vertices in $(a,\phi_n)$ in polar coordinates can be expressed as
\begin{equation}
    \frac{\hat{J}_n-J_n}{J_n}=\frac{\sum_{q=-\infty}^\infty U_q \upsilon_q \, e^{-\mathrm{j}q\phi_n}}{\sum_{q=-\infty}^\infty U_q \, e^{-\mathrm{j}q\phi_n}}\,,
\end{equation}
where $U_q^{\text{TM}} = 2 \mathrm{j}^{-q}/(\pi \eta k a  H_q^{(2)}(k a))$ and $U_q^{\text{TE}} = 2 \mathrm{j}^{-q}/(\pi \eta k a  H_q^{(2)'}(k a))$. As already shown in \cite{warnick2008numerical}, for the EFIE
\begin{equation}
    \upsilon_q^{\text{TM/TE-EFIE}}=\frac{T_{-q}(1-F_q)-E_q^{A,\op{S}^k/\op{N}^k}}{1+E_q^{\op{S}^k/\op{N}^k}}\,.
\end{equation}
Similarily, for the MFIE
\begin{equation}
    \upsilon_q^{\text{MFIE}}=\frac{T_{-q}(1-F_q)-E_q^{A,\text{MFIO}^k}}{1+E_q^{\text{MFIO}^k}}\,.
\end{equation}
In the CCFIE case instead our derivations, omitted here due to space constraints,  show that $\upsilon_q$ is given by the weighted average
\begin{equation}
    \upsilon_q^{\text{CCFIE}}=\frac{\hat{\lambda}_q^{\text{CEFIO}}\upsilon_q^{\text{EFIE}}+\hat{\lambda}_q^{\text{CMFIO}}\upsilon_q^{\text{MFIE}}}{\hat{\lambda}_q^{\text{CEFIO}}+\hat{\lambda}_q^{\text{CMFIO}}}\,.
\end{equation}

Different measures of the current relative error are available and significant for diverse purposes. We consider here the $L^2-$measure,
\begin{equation}
    r_{L^2(\Gamma)} \coloneqq \left(\frac{\sum_{q=-\infty}^\infty|U_q \upsilon_q|^2}{\sum_{q=-\infty}^\infty|U_q|^2} \right)^{1/2}\,,
\end{equation}
the measure in the standard norm of the current space $H^s(\Gamma)$, with $s=\mp 1/2$ for the TM/TE formulations,
\begin{equation}
    r_{H^s(\Gamma)} \coloneqq\left(\frac{\sum_{q=-\infty}^\infty|U_q \upsilon_q|^2\, (1+q^2)^s}{\sum_{q=-\infty}^\infty|U_q|^2\, (1+q^2)^s} \right)^{1/2}\,,
\end{equation}
and the measure in a different norm in $H^s(\Gamma)$,
\begin{equation}
    r_{H_k^s(\Gamma)} \coloneqq\left(\frac{\sum_{q=-\infty}^\infty|U_q \upsilon_q|^2\, (k^2+q^2)^s}{\sum_{q=-\infty}^\infty|U_q|^2\, (k^2+q^2)^s} \right)^{1/2}\,,
\end{equation}
commonly used in high-frequency scattering applications \cite{chandler-wilde2015wavenumberexplicit}.

Following the spectral analysis in Section \ref{sec:spectralerror}, one can show that, for indices $q\simeq (ka)$, $|\upsilon_q^{\text{TM-EFIE}}|$ and $|\upsilon_q^{\text{TM/TE-MFIE}}|$ have a constant behavior in frequency. On the contrary, the increasing behavior as $k^{1/3}$ of the aliasing spectral error of the hypersingular operator translates into an increase of $|\upsilon_q^{\text{TE-EFIE}}|$ at the same rate. Hence, the relative current error in the three measures considered does not increase in frequency for the TM-EFIE, TM/TE-MFIE and TM-CCFIE, while it does for the TE-EFIE. The Calder\'{o}n preconditioning and combination of equations in the TE-CCFIE on the other hand significantly attenuates the error increase due to the increase of $|\upsilon_q^{\text{TE-EFIE}}|$.

\section{Numerical Results}
\label{sec:numerical}
\begin{figure}
\centerline{\includegraphics[width=1\columnwidth]{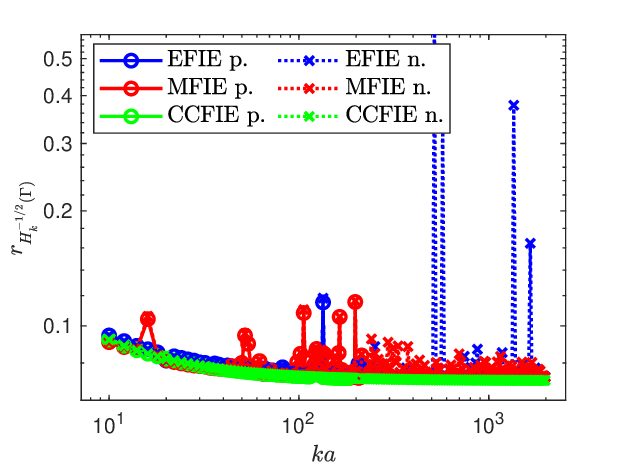}}
\caption{Frequency dependency of $r_{H^{s}_k(\Gamma)}$ of the TM formulations: comparison between predicted (p.) values and numerical (n.) results from BEM.}
\label{fig:jHskTM}
\end{figure}

\begin{figure}
\centerline{\includegraphics[width=1\columnwidth]{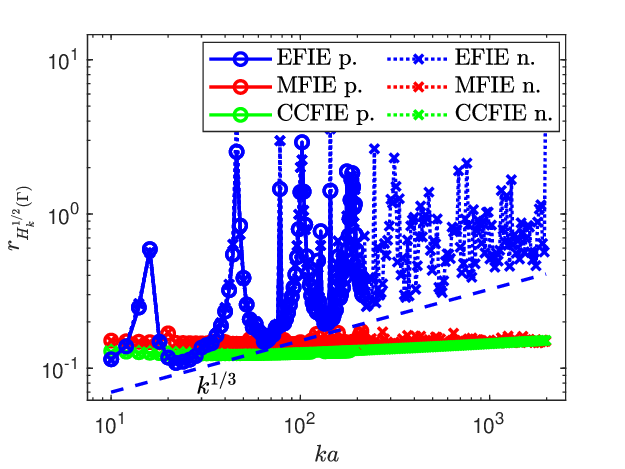}}
\caption{Frequency dependency of $r_{H^{s}_k(\Gamma)}$ of the TE formulations: comparison between predicted (p.) values and numerical (n.) results from BEM.}
\label{fig:jHskTE}
\end{figure}

To showcase the validity of the theoretical results presented in Section~\ref{sec:novelty}, we estimated the current error for increasing frequency, both out of the application of the formulae above and from the solution of our BEM solver.
The geometry has been discretized at length $h$ approximately equal to the wavelength over four. Testing and source basis functions employed have polynomial order \num{1}. Figures \ref{fig:jHskTM} and \ref{fig:jHskTE} show the frequency dependency of $r_{H_k^{s}(\Gamma)}$ for TM and TE formulations. The circles, representing values from formulae, and crosses, representing values from the BEM solver, are in good agreement. 
In the EFIE and MFIE case the error increases sharply in correspondence of resonance frequencies, which are approximately the same for the TM-EFIE and the TE-MFIE and for the TM-MFIE and the TE-EFIE, while the CCFIE is immune from spurious resonances.
Finally, we notice the increase as $k^{1/3}$ of $r_{H_k^{s}(\Gamma)}$ for the TE-EFIE current away from resonances, resulting from the same order increase of $|\upsilon_q^{\text{TE-EFIE}}|$ in the transition region.


\section{Conclusion}
We analyzed the effects of the discretization of boundary integral operators on the spectra of their discrete counterparts. We proceeded in studying the consequences on the achievable accuracy of some of the most common equations in computational electromagnetics.

\section*{Acknowledgements}
This work was supported by the European Innovation Council (EIC) through the European Union’s Horizon Europe research Programme under Grant 101046748 (Project CEREBRO) and by the European Union – Next Generation EU within the PNRR project ``Multiscale modeling and Engineering Applications'' of the Italian National Center for HPC, Big Data and Quantum Computing (Spoke 6) – PNRR M4C2, Investimento 1.4 - Avviso n. 3138 del 16/12/2021 - CN00000013 National Centre for HPC, Big Data and Quantum Computing (HPC) - CUP E13C22000990001.

{ \printbibliography}

\end{document}